\documentclass[prl,aps,twocolumn,showpacs]{revtex4-1} 
\usepackage{graphicx}
\begin{document}

\def\ii{\'\i}
\def \st {\widetilde \sigma}
\def \sd {\sigma ^{\dag}}
\def \pt {\widetilde\pi}
\def \pd {\pi ^{\dag}}
\def \fn {$\Phi_{NLM}$}
\def \fav {$\Phi_{A^{'}}$}
\def \fx {$\Phi _{x}$}
\def \fa {$\Phi _{A}$}
\def \np{n_{\pi}}
\def \f7 {$f_{7/2}$}
\def \si {$\sigma$}
\def \oa {$^{16}O + \alpha$ \ }
\def \ca {$^{12}C + \alpha$ \ }
\def \cam {$^{12}C + \alpha$}
\def \co {$^{12}C + \ ^{16}O$ \ }
\def \com {$^{12}C + \ ^{16}O$}
\def \ma {$^{24}Mg + \alpha$}
\def \mam {$^{24}Mg + \alpha$ \ }
\def \cc {$^{12}C + \ ^{12}C$ \ }
\def \ust {$U^{ST}(4)$ \ }
\def \ustc {$U^{ST}_C(4)$ \ }

\pagebreak
\title{On the intersection of the shell, collective and cluster models of atomic nuclei II:
Symmetry-breaking and large deformations}
\author{J.~Cseh}
\affiliation{Institute for Nuclear Research, Hungarian Academy of Sciences, Debrecen, Pf. 51, Hungary-4001}
\author{J.~Darai}
\affiliation{Department of Experimental Physics, University of Debrecen, Debrecen, Pf. 105, Hungary-4010}
\date{\today}

\begin{abstract}
We discuss the  role of the broken symmetries in the connection of the  shell, collective and cluster models.
The cluster-shell competition is described in terms of cold quantum phases. 
Stable quasi-dynamical U(3) symmetry is found for specific  large deformations for a Nilsson-type Hamiltonian.

\end{abstract}

\maketitle

\section{Introduction}

The connection of the shell, collective
\cite{elli58}
and cluster models 
\cite{wika58,babo58},
found in terms of the SU(3) symmetry in 1958, 
was based on a single-shell problem, described in spherical basis, related to simple symmetries.
In a pevious paper
\cite{prev}
the extension to multi-major shells was addressed, and the 
U(3) $\otimes$ U(3) $\supset$ U(3)
dynamical symmetry was found as the common intersection of the three models.
Here we consider the role of more general symmetries and the case of large deformation.

So far two kinds of the SU(3) symmetry were applied in this respect.
i) The exact symmetry, in which case 
both  the Hamiltonian is symmetric, i.e. 
it is an SU(3) scalar, and  its eigenvectors are symmetric, i.e.  they transform according to
irreducible representations (irreps). 
ii) The dynamical SU(3) symmetry (sometimes called dynamically broken symmetry),
when the eigenvectors are still symmetric, but the interactions are not.
This kind of special breaking is achieved by incorporating 
an interaction  which is expressed in terms of the invariant operator of the
SO(3) subgroup, in addition to the   SU(3)  scalar part.

The relation between the shell and collective models, established by Elliott
\cite{elli58}
was based on the SU(3) $\supset$ SO(3) dynamical symmetry.
The Wildermuth-connection between the shell and cluster models 
was originally based on harmonic oscillator Hamiltonians of exact SU(3)
symmetry, but it turns out to be valid also for the dynamical symmetry, 
as discussed e.g. in
\cite{musyk}.

Here we consider the more general quasi-dynamical SU(3) symmetry
\cite{quasi}
which turns out to be important, too.
Therefore, in the next section we discuss very briefly the hierarchy of symmetries
(or symmetry-breaking), relevant for the connection of the fundamental structure 
models. 
We also describe the cluster-shell competition in terms of quantum phases,
being closely related to symmetries.
Then we investigate the case of large defomations, with special emphasis on
the symmetries of the super- and hyperdeformation, in the presence of realistic
(Nilsson-type) interactions. Before the concluding part we devote a short section
to the comparison of different kinds of extensions of Elliott's original SU(3)
symmetry, which are relevant from the present viewpoint. The $^{20}$Ne nucleus is applied
for illustrative purposes, just like in 
\cite{prev}.

\section{Symmetries and symmetry-breaking}

\subsection{Hierarchy}

The quasi-dynamical (or effective) U(3) symmetry 
\cite{quasi} is a
generalization of the concept of the  U(3) dynamical symmetry
 in the following sense. 
The Hamiltonian breaks the symmetry in such a way that the U(3) 
quantum numbers are not valid for its eigenvectors  (contrary to the case
of the real U(3) dynamical symmetry). In other words neither the
operator is symmetric,
% (i.e. it is not a U(3) scalar), 
nor its eigenvectors 
%(i.e. they do not transform according to  irreducible representations)
\cite{varna}.
(For comparison with the exact and dynamical symmetries  see Table I.) 
Yet, the symmetry is present in some sense, and it
may survive even for strong
symmetry-breaking interactions
\cite{quasi}. 
Then the energy eigenstates are:
\begin{equation}
\psi_{\alpha K J M} = 
\Sigma _{ \xi \lambda \mu}
C_{\alpha \xi \lambda \mu K}
\phi_{\xi \lambda \mu K J M} ,
\end{equation} 
where $\phi_{\xi \lambda \mu K J M}$ is a basis vector for an 
SU(3) irreducible representation,  
$\xi$ and $\alpha$ are additional quantum numbers needed 
to specify the wavefunction
\cite{jarrio}.
The 
$C_{\alpha \xi \lambda \mu K}$
coefficients of the linear combination
are independent of $J M$, i.e. within a band the contribution of different SU(3) basis states
is the same. 
(This situation is called adiabatic approximation.)
When calculating the matrix elements of the SU(3) generators between these states
the result may approximate the matrix elements of an exact representation.
In such a case we speak about an approximate embedded representation, and related 
to it, about an approximate quasi-dynamical or effective SU(3) symmetry.

\begin{table}
%\begin{table*}
\caption{Different  SU(3) symmetries in nuclear stucture models.
The signs '+' and '-' indicate if the Hamiltonian and its eigenvectors
are symmetric or not. The operator is said to be symmetric, if it is
a scalar, while a set of eigenvectors is symmetric if they transform
according to  irreducible representations.
}
\begin{tabular}{|c|c|c|c|}
\hline
\hline
symmetry&Hamilt.&eigenvect.&model\\
\hline
\hline
exact&+&+&harmonic oscillator\\
\hline
dynamical (br.)&-&+&Elliott, IBM, SACM,...\\
\hline
quasi-dynam.&-&-&shell, quantum phase,...\\
\hline
\hline
\end{tabular}

%\end{table*}
\end{table}

An asymptotic Nilsson-state serves as an intrinsic state for the quasi-dynamical
SU(3) representation. Thus the effective quantum numbers are determined by the
Nilsson-states in the regime of large deformation
\cite{jarrio}.
When the deformation is not large enough, then we can expand the
Nilsson-states in the asymptotic basis, and calculate the effective quantum
numbers based on this expansion
\cite{epj}.

The concept of effective U(3) symmetry is applicable also 
for the case when the simple leading
representation approximation is valid, and then the real and effective 
U(3) quantum numbers usually
coincide
\cite{epj}.
The quasi-dynamical symmetry appears in the investigation of both quantum phases
and the large deformation.

\subsection{Phases}
%\subsection{Diagrams}
%\subsection{Locations}

Symmetry-adapted models are especially suitable for the studies
of the phases and phase-transitions in finite quantum systems
\cite{phase}.
The usual scenario is to consider an algebraic model
with a well-defined model space
and with interactions which are varied continuously. 
The model has limiting cases, i.e. dynamical symmetries.
When a dynamical symmetry holds, the eigenvalue-problem has an analytical
solution. The general Hamiltonian, however, which has
contributions from interactions with different dynamical symmetries, has to be 
diagonalized numerically. The relative weight of the dynamically symmetric 
interactions serves as a control parameter, and it defines the 
phase-diagram of the system. When there are more than two dynamical symmetries,
more than one control parameters appear.

In the limit of large particle number 
phase-transitions are seen in the sense that the
derivative of the energy-minimum, as a function of the control-parameter, 
is discontinuous. The order of the derivative, showing the
discontinuity, gives the order of the phase-transition. Thus the 
phase-transition is investigated quantitatively, like in the
thermodynamics. A phase is defined as a region of the phase diagram between the
endpoint of the dynamical symmetry and the transition point. It is also
conjectured 
\cite{rowephase} 
that such a quantum phase is characterised by a quasi-dynamical
symmetry. Therefore, although the real dynamical symmetry is valid only at a 
single point of the phase-diagram, the more general quasi-dynamical symmetry 
may survive, and in several cases does survive
\cite{rowephase,hjp}, 
in a finite volume of the phase diagram. 
If this conjecture really turns out to be true, then 
the situation is similar to Landau's theory: different
phases are determined by different (quasi-dynamical) symmetries, and phase
transitions correspond to a change of the symmetry. 

In the case of the finite particle number the discontinuities are smoothed out, 
as the consequence of the finite size effect, but still remarkable
changes can be detected in the behaviour of the corresponding functions.

The semimicroscopic algebraic cluster model (SACM)
\cite{sacm}
 of a binary cluster system has three dynamical symmetries, see
(2).
%Eq.
%\ref{}
%(17).
%(\ref{3dysy}).
Two of them come from the vibron model of the relative motion:
U$_R$(3) corresponds to shell-like clusterization, or in the language
of the collective motion to a soft vibrator, while O$_R$(4) represents a rigid 
molecule-like rotator. The third symmetry, U$_R$(3)$\supset$SO$_R$(3) 
corresponds to a situation,
when the coupling between the relative motion and internal degrees of freedom
of the clusters is weak. Therefore, the phase diagram of a binary system is
two-dimensional, and can be illustrated by a triangle
\cite{varna9}.

Schematic calculations show that 
there is a second order phase transition
(in between the shell-like (U$_R$(3)) and rigid
molecule-like (O$_R$(4)) clusterizations) 
when interaction terms up to second order are included, but also a first order
transition shows up, if third order interactions are involved
\cite{fraser}.

\begin{widetext}
\begin{eqnarray}
U_{C_1}(3) \ \ \otimes U_{C_2}(3) \ \ \otimes \ \ U_R(4) \
\supset  U_C(3)   \ \ \otimes \ \ U_R(3) \ \
 \supset
 \ \ U(3) \ \supset SU(3)\ \ \ \  \supset SO(3) \supset SO(2)
 \nonumber \\
 U_{C_1}(3) \ \ \otimes U_{C_2}(3) \ \ \otimes \ \ U_R(4) \
 \supset
U_C(3) \ \ \otimes \ \ O_R(4) \ 
\supset SO_C(3)
 \otimes SO_R(3) \ \
 \supset SO(3) \supset SO(2)
% \nonumber \\
 \\
 \label{3dysy}
 U_{C_1}(3) \ \ \otimes U_{C_2}(3) \ \ \otimes \ \ U_R(4) \
 \supset
U_C(3) \ \ \otimes \ \ U_R(3) \ 
\supset SO_C(3)
 \otimes SO_R(3) \ \
 \supset SO(3) \supset SO(2)
 \nonumber
% \\
\label{3dysy}
\end{eqnarray}
\end{widetext}

\begin{figure}
\includegraphics[height=7.cm,angle=0.]{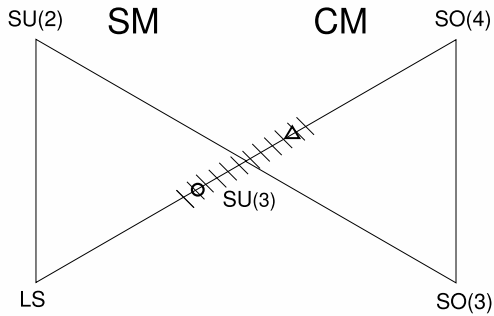}
\caption{ Joint phase diagram of the shell and cluster models, and the location of the ground-band of
the $^{20}$Ne nucleus, from shell (circle), cluster (triangle) and quasicluster (diffuse) model calculations. 
%For more details see the text.
\label{fig:csokinyaki}}
\end{figure}

A triangle-like phase diagram has been proposed for the shell model, too
\cite{pvi},
which in addition to the SU(3) and SU(2) symmetries
has the independent-particle model as the third corner.
The limiting cases correspond to the situations, in which  the
quadrupole-quadrupole,  pairing (both of them are two-body)
interactions, and the single-particle energies dominate, respectively.
(When a single shell calculation is performed then 
the spin-orbit interaction may be the most relevant
part of the single-particle contribution.)
The two phase diagrams match each other at the SU(3) corner,
as shown in Figure 
\ref{fig:csokinyaki}.
The real nuclear systems can be allocated to this diagram. 
The control parameters  measure the distance
from the dynamical symmetries, as mentioned before.
%Another approximate method, which is applicable not only for algebraic
%models is based on the SU(3) expansion of the wavefunction.
%Since the SU(3) basis states represent 
%the intersection of the shell and cluster models,
%their distribution may indicate the closeness to the matching point.
%In the extreme case of a single SU(3) basis state, the overlap is 100 percent,
%and the system sits right at the intersection of the two phase diagram. 
%In general,  the weight of the leading representation shows the overlap, 
%{\textit i.e.} how much the state is a shell-like cluster state. 

Here we refer to the results of three different calculations on the ground-band of the
$^{20}$Ne nucleus. Vargas et al
\cite{vargas}
performed a full shell calculation 
for the 4 nucleons outside the core
(accounting for many other states in addition to the ground band, of course).
They applied  SU(3) basis, and the Hamiltonian included, in addition to the harmonic oscillator 
potential, quadrupole-quadrupole, pairing interactions of the alike nucleons, 
and spin-orbit force (as well as some rotor-like terms to fine tune the moment of
inertia and the position of different $K$-bands). 
For the purpose of illustration the parameters of their Hamiltonian
%(in arbitrary units) 
can be rewritten into the form
\begin{equation}
H = yz H_{SU3} + (1-z)H_{LS} + (1-y)zH_{SU2}
\label{vargas}
\end{equation} 
where the control parameters $y,z$ take values between 0 and 1,
in such a way that 
$z=1, \ y=1$ corresponds to the SU(3) dynamical symmetry,
$z=1, \ y=0$ corresponds to the SU(2) dynamical symmetry, and
$z=0$ refers to the limit of the large $LS$ interaction.
(More specifically we took the parametrization, as follows:
$
H = \Sigma_i  a_i H_{iSU3} + bH_{LS} + cH_{SU2} ,
$
$ \vert a \vert =  \Sigma_i  \vert a_i \vert$,
$y= \vert a \vert / (\vert a \vert + \vert c \vert) $,
$z=  (\vert a \vert + \vert c \vert)/ (\vert b \vert + \vert a \vert + \vert c \vert) $,
where $\vert w \vert$ indicates  the absolute value of $w$.)
The location of the system on the shell model diagram is 
$y=0.99, \ z= 0.77$.

Yepez-Martinez et al  have carried out a similar calculation on the
cluster-side of the phase diagram
\cite{huitz20ne}. They applied the 
semimicroscopic algebraic cluster model
\cite{sacm}
with a Hamiltonian:
\begin{equation}
H = xv H_{SU3} + (1-x)vH_{SO4} + (1-v)H_{SO3} ,
\label{huitz}
\end{equation} 
therefore, 
$v=1, \ x=1$ corresponds to the SU(3) dynamical symmetry,
$v=1, \ x=0$  shows the SO(4) dynamical symmetry, and
$v=0$ refers to  the SO(3) limit.
They have found
$v=1.00, \  x=0.78$.
(In this calculation several other cluster bands were obtained, too.)

Itagaki et al applied the antisymmetrized quasi-cluster model
\cite{itag}
for the description of the ground band.
This model can take a direct route from the rigid molecule-like
(SO(4)) clusterization via the shell-like cluster limit (SU(3))
to the $jj$-coupled shell model dominated by a strong
$LS$ interaction.
Thus, it is especially illuminative from the viewpoint of the
cluster-shell competition. It does not have, however, a well-defined 
algebraic structure, therefore, the control parameters can not be introduced
in terms of the relative weights of interactions with different dynamical
symmetries. It  has two parameters to characterize the situation,
but they are parameters of the wavefunction. One of them ($R$) refers to
the distance of the (quasi) clusters, the other ($\Lambda$) is related to the strength
of the spin-orbit force. A qualitative correspondance can be found
between these parameters and the ones mentioned before in relation
with the algebraic shell and cluster models. In short: the
small $R$ and small $\Lambda$ is located near the matching point of the
two diagrams (SU(3) dynamical symmetry), increasing $R$ takes towards
the rigid molecule-like (SO(4)) corner, while inreasing $\Lambda$
moves to the large $LS$ limit. The result of this study showed that
the experimental situation corresponds to the shell like clusterization, i.e.
close to the crossing point between the shell and cluster model.

It is remarkable that  three different model calculations
have similar conclusion  on the closeness of the ground-band of $^{20}$Ne
to the matching point between the shell and cluster models.

\section{Large deformations}

The  SU(3) connection from 1958 is based on the symmetry of the spherical
shell model. In light of the fact that today very largely deformed states
are known experimentally, it is an important question, what happens
to this symmetry with increasing deformation. The superdeformed states
e.g. represent a situation which is close to the axially symmetric spheroid
with ratios of main axes of 2:1:1, the hyperdeformed state corresponds to
3:1:1.

\subsection{Deformed harmonic oscillator}

In \cite{rodr} it was shown that the symmetry algebra of the anisotropic
harmonic oscillator is SU(3),  whenever its frequencies are commensurate, i.e.
expressed as ratios of integer numbers. As a special case it includes the spherical
oscillator, as well as the superdeformed or hyperdeformed shapes.
More details, concerning the axially symmetric case with 2:1, 3:1, and 3:2
ratios are discussed in 
\cite{suar97}, and  the triaxially deformed oscillator in
\cite{suar09}.
(For previous works on this problem we refer to the citations in these
papers.)

\begin{figure}
\includegraphics[height=6.0cm,angle=0.]{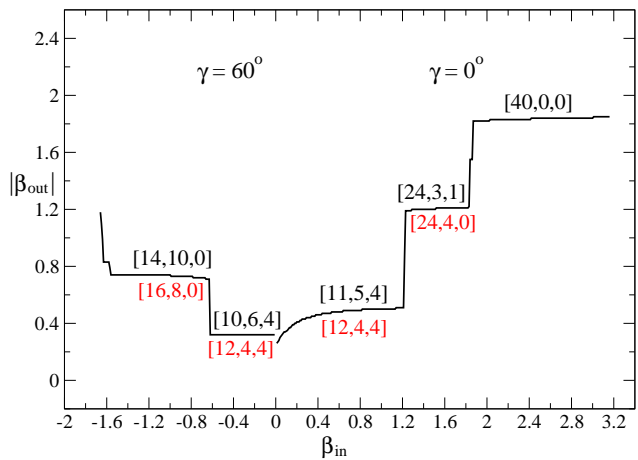}
\caption{ 
Shape isomers of the  $^{20}$Ne nucleus from Nilsson model and quasi-dynamical SU(3) symmetry calculations.
The effective U(3) quantum numbers are shown above the curve, while the ones belonging to simple
harmonic oscillator configurations are indicated below (when they are different).
\label{fig:lepcsos}}
\end{figure}

The connection between  the anisotropic harmonic oscillator and clusterization
have been discussed in 
\cite{ra88,nado92,cssc92,freer}.

\subsection{Realistic interactions}

In considering  realistic interactions the Nilsson model plays an important role:
it gives the single particle orbits of a Hamiltonian with
spin-orbit, as well as $l^2$ terms
\cite{bera}:
\begin{equation}
H =  \hbar \omega N  + \hbar \omega r^2 \beta  Y_{20} ( \theta , \phi)
%    + C {\bf l} {\bf s} + D {\bf l}^2 .
    + C  l s + D  l^2 .
\end{equation}

Soon after the experimental discovery of the superdeformed 
states, Sugawara-Tanabe et al have realised that the $L-S$ coupling recovers
for the superdeformed shape, first in a simple
\cite{suar93}, then in more realistic Nilsson model calculations
\cite{suar95}. 
In particular they found that the $L-S$ coupled spherical wavefunction components
became more than 85\% of the total wavefunction.
The reason for this is the dominant role of the
quadrupole interaction due to the large deformation. This phenomenon provides an
explanation for the appearance of the parity doublet levels.

In what follows we investigate the survival (or appearance) of the SU(3) 
symmetry systematically as a function of the quadrupole deformation
(including triaxiality). We do so in terms of the Nilsson-model
combined with the concept of  the quasi-dynamical symmetry, discussed
in the previuos section.
We obtain the shape isomers from a  selfconsistent calculation concerning the quadrupole
deformation. In particular:
one varies systematically  the  parameters ($\beta_{in} , \gamma_{in}$),
as an input for the Nilsson-model. The calculations provide us with  the effective U(3) quantum numbers,
which can be translated into  the ($\beta_{out} , \gamma_{out}$) quadrupole deformation, since
they are  uniquely related to each other
\cite{bg}.
Then one can check if the selfconsistency is satisfied, as well as the question whether or not
the result is stable with respect to the (small) changes of the input values.
This method for the determination of the shape isomers is an alternative to the standard 
energy-minimum calculation  and has been shown to be effective for a range of light 
nuclei~\cite{ar36,ni56,si28}.

The result of the Nilsson model + quasi-dynamical SU(3) calculation 
for $^{20}$Ne is shown in Figure~\ref{fig:lepcsos},
for $\gamma_{in} = 60^o$ and $\gamma_{in} = 0^o$. In this figure, the horizontal plateaus rather 
than the minima correspond to stable shapes. (The calculations with $\gamma$ in between $0^o$ and  $60^o$
do not show other shape isomers. In fact those of  $10^o$,  $20^o$ and  $30^o$  are very similar
to the right side of Figure~\ref{fig:lepcsos}, while the  $40^o$ and  $50^o$ resemble to its left side.)

%Similar calculations have been performed for intermediate values of $\gamma$ in steps of $5^o$; 
%the results for $\gamma_{in} \le 30^o$ are broadly similar to those for $\gamma_{in} = 0^o$, 
%and calculations for $\gamma_{in} > 30^o$ are similar to those for $\gamma_{in} = 60^o$. 

\begin{table}
\caption{Shape isomers in the $^{20}{\rm Ne}$ nucleus from the Nilsson model and quasi-dynamical 
 SU(3) calculation. The states are the ground (GS), superdeformed (SD), hyperdeformed (HD),
and linear $\alpha$-chain (AC) states.
`e' stands for effective U(3) quantum numbers, `h' indicates the states 
corresponding to simple harmonic oscillator configurations. 
The triaxiality, $\gamma$, is given in degrees. The ratio of the main axes is denoted by a:b:c.
The '+' sign in the column BB indicates that  the Bloch-Brink $\alpha$-cluster calculation 
\cite{1d,2d,3d}
gave a state
with the same U(3) symmetry. LL refers to the calculation by Leander and  Larsson
\cite{lale}, 
citing their value for $\gamma$.
\label{tab:shis}}
\begin{center}
\begin{tabular}{|c|c|c|c|c|c|c|c|c|}
\hline
\multicolumn{1}{|c|}{State}
& \multicolumn{1}{|c|}{Q.no.}
& \multicolumn{1}{|c|}{$\hbar\omega$}
& \multicolumn{1}{|c|}{U(3)}
& \multicolumn{1}{|c|}{$\beta$}
& \multicolumn{1}{|c|}{$\gamma$}
& \multicolumn{1}{|c|}{a:b:c}
& \multicolumn{1}{|c|}{BB}
%& \multicolumn{1}{|c|}{$\cal{J}$$^{(1)}$}\\
& \multicolumn{1}{|c|}{LL}\\
\hline
\hline
GS&e&0&[11,5,4]&0.43&7.6&1.5:1.1:1&&0\\
&h&0&[12,4,4]&0.52&0.0&1.6:1.0:1&+& \\
\hline
SD&e&4&[14,10,0]&0.75&43.9&2.4:2.0:1&&50\\
&h&4&[16,8,0]&0.84&30.0&2.6:1.8:1&+& \\
\hline
HD&e&8&[24,3,1]&1.24&4.5&3.1:1.2:1&&\\
&h&8&[24,4,0]&1.25&8.9&3.4:1.4:1&+& \\
\hline
AC&e&20&[40,0,0]&1.85&0.0&5.0:1.0:1&&0\\
&h&20&[40,0,0]&1.85&0.0&5.0:1.0:1&+& \\
\hline
\end{tabular}
\end{center}
\end{table}

Table \ref{tab:shis} compares the results of the present calculations with those of the 
earlier determination of shape isomers 
using Nilsson model potential energy surfaces~\cite{lale}, and the Bloch-Brink alpha-cluster 
model~\cite{1d,2d,3d}. 
Leander and Larsson list three shape isomers \cite{lale}, each of which is very close to our 
present ones. The alpha-cluster studies found all the four shape isomers, we see here.
Their U(3) symmetries are in coincidence with those of the simple harmonic oscillator
configurations approximating the effective quantum numbers. It is remarkable that
the instability of the shape in the ground state region, shown by the  curvature in
Fig.~\ref{fig:lepcsos}, 
 is observed also in the Bloch-Brink calculations
\cite{3d};
they mention two close-lying states with very similar configurations.

\subsection{Connection to clusterization}

The U(3) connection 
\cite{elli58,babo58}
between the collective, shell and cluster models works
well in case i) the U(3) symmetry is approximatelly valid, and 
ii) the relation between the cluster and shell model wavefunctions is simple.
This latter condition means that  the expansion of the cluster U(3) state in terms 
of shell basis reduces to a  few terms.

As discussed in the previous subsection the U(3) symmetry recovers for
the superdeformed, hyperdeformed, etc shapes, in spite of the important
role of the symmetry-breaking  spin-orbit  interaction. 
The second condition (on the simple shell model expansion) turns out to
be valid also for several shape isomers.  E.g. in case of  the $^{20}$Ne
nucleus each of the 4 shape isomers (of the U(3) symmetry: [12,4,4], [16,8,0],
[24,4,0], [40,0,0]) has  single multiplicity in the shell model basis.
Therefore, if a cluster state with the same U(3) symmetry is allowed
it is identical with the shell state. This is a consequence of the fact that
the cluster state can be expanded in terms of shell basis, and basis states
of different U(3) irreps are orthogonal to each other. (Both
the shell and cluster states are  normalized.)

In general: the U(3) selection rule for determining the allowed clusterization
is:
\begin{equation}
[n_1, n_2, n_3] = [n^{C_1}_1, n^{C_1}_2, n^{C_1}_3] \otimes
 [n^{C_2}_1, n^{C_2}_2, n^{C_2}_3] \otimes  [n^{R}, 0, 0],
\end{equation}
where 
$[n_1, n_2, n_3]$ is the U(3) symmetry of the shell model state,
$ [n^{C_i}_1, n^{C_i}_2, n^{C_i}_3]$
is that of the $i$th cluster, and 
 $[n^{R}, 0, 0]$ stands for the relative motion.
When the shell model irrep matches with one of the results of the triple product
of the right hand side, the cluster configuration is allowed.
In addition to the U(3) selection, there is another simple prescription 
by Harvey
\cite{harvey}
for the determination of the allowed clusterizations.
It also applies harmonic oscillator basis, so the two requirements are somewhat similar.
Nevertheless, their physical content is not the same; in some sense
they are complementary to each other.
Therefore, the best way is to apply them in a combined way
\cite{binary,ternary,dubna}.
 (Their relation is discussed more in detail in
\cite{fusion}.)
When a cluster configuration is forbidden, one can characterize its forbiddenness
quantitatively
\cite{forbid}.

The energetic preference represents a complementary viewpoint for
the selection of clusterization. We usually incorporate it in two different ways:
i) by applying simple binding-energy  arguments
\cite{FBS2000},
and ii) by performing double-fol\-ding
calculations, according to the dinuclear system model
\cite{dns,PLB2002}. 

The alpha-like (N=Z=even) cluster configurations  are energetically preferred, in general.
When considering binary clusterizations with both clusters in their intrinsic ground state,
then  the GS state of $^{20}$Ne allows both $^{16}$O+$^{4}$He, and 
$^{12}$C+$^{8}$Be clusterizations. The latter configuration is present also
in the SD and HD states, while the linear alpha-chain can not be built up from
two ground-state clusters.
Due to the single shell model multiplicity of these shape isomers, one can say that
to the extent the U(3) symmetry is valid, these states can be considered, as the cluster
configurations mentioned before.

\section{Extension of Elliott's  SU(3) symmetry}

\begin{figure}
\includegraphics[height=6.5cm,angle=0.]{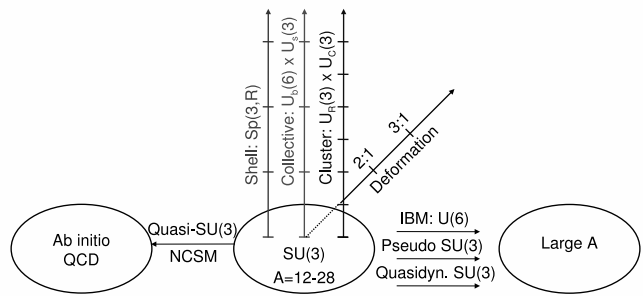}
\caption{ 
Extension of Elliott's SU(3) symmetry in different directions.
\label{fig:elliott}}
\end{figure}

In Fig.
\ref{fig:elliott}
we illustrate some generalizations of Elliott's SU(3) symmetry into different directions.
In its original form it proved to be effective for light nuclei of the $p$ and $sd$ shell,
for a single major shell problem.
The vertical extension along the shell-excitation has been discussed in detail in 
\cite{prev}.
 
Several approaches have been invented in order to export the elegant and powerful 
technique of the group theory to the medium and heavy nuclei. The interacting boson model 
\cite{ibm} uses a method (by introducing bosons of coupled valence nucleons) which is applicable 
for a single (or a few) major shell.
This model has a U(6) group structure, and one of its dynamical symmetries: 
U(6)$\supset$SU(3)$\supset$SO(3)
describes the rotational spectra of deformed nuclei.
The  IBM model has been applied to heavy nuclei in a very wide range.

The pseudo-SU(3) symmetry associates the SU(3)  irrep of the $(n-1)$-th major shell to
a subset of the states in the $n$-th major shell. It appears due to a special ratio of
the $ls$ and $l^2$ interactions
\cite{pseudo}.  
Thus the nucleon states are divided into two
categories, and one of them carries the SU(3) symmetry. This approach  can treat several
major shells in a similar manner.

The quasi-dynamical SU(3) symmetry, as discussed beforehand, builds up from the
contribution of many nucleons in a "democratic" way, i.e. no distinction is
made between the  single particle levels. It turns out that this symmetry may survive
in the presence of different symmetry-breaking interactions, like e.g. spin-orbit and pairing
\cite{rowedr}.

The extension towards the very light nuclei is related to  the quasi-SU(3) symmetry
(not to be mixed up with the quasi-dynamical symmetry, mentioned before)
\cite{zuke95}
and no-core shell model (NCSM). 
The quasi-SU(3) is a symmetry of the shell model, and in the $LS$ coupled proton-neutron 
formalism it results in an efficient truncation scheme: only the low spin components 
and SU(3) basis states of large deformation give important contribution
\cite{varg98,vargas}. 
It is interesting that it turned out to be effective also with realistic nucleon-nucleon interactions
in ab intio calculations
\cite{dytr07}. 

The generalization to deformed basis and large deformation has been discussed in the previous
section.

\section{Summary}

In the present and the previous paper we have discussed the extension of the SU(3)
connection between the shell, collective and cluster models. This relation was established
in 1958
\cite{elli58,wika58,babo58}
for a single major shell, by applying spherical basis of the exact or dynamically broken 
symmetry. 
In
\cite{prev}
we considered the vertical extension, i.e. the incorporation of major shell excitations.
Here we discussed further generalization along the symmetry-breaking and large deformation.

The cluster-shell competition or coexistence has been interpreted in terms of the joint phase 
diagram of the shell and cluster models. Three different (shell, cluster and quasi-cluster)
calculations indicate the position of the ground-band of $^{20}$Ne very close to the SU(3)
matching point of the two models.

Concerning the large deformations the quasi-dynamical SU(3) symmetry is found to be
stabil for Nilsson-type interactions for several shapes with commensurable ratios of the
main axes.

All these considerations, as well as many others, are based on the extension of Elliott's
SU(3) symmetry. In Figure 3 we summarized some of its generalization along
different directions: excitation energy, mass number and deformation.

\section{Acknowledgement}
This work was supported by the OTKA (Grant No K106035), 
as well as by the MTA-BAS (SNK 6/2013) and  MTA-JSPS  (No. 119) bilateral projects.
Inspiring discussions with 
Professors  M. Freer,  N. Itagaki, and M. Ploszajczak are
gratefully acknowledged.

\end{document}